\documentclass[multphys,vecphys]{svmult}

\usepackage{makeidx}         
\usepackage{graphicx}        
\usepackage{multicol}        
\usepackage[bottom]{footmisc}

\makeindex             

\begin{document}

\title*{3D spectroscopy as a tool for investigation of the BLR of lensed 
QSOs
}
\titlerunning{3D spectroscopy as a tool for investigation of the BLR}
\author{Luka \v C. Popovi\'c
}
\institute{Astronomical Observatory, Volgina 7, 11160 Belgrade, Serbia
\texttt{lpopovic@aob.bg.ac.yu}\footnote{The work is supported by the 
Ministry
of Science and Environment
Protection of Serbia through the project ``Astrophysical
Spectroscopy of Extragalactic Objects'' and by
Alexander von Humboldt Foundation through "R\"{u}ckkehrstipendium".}
}
%
%
\maketitle

\begin{abstract}
Selective amplification of the line and
continuum source by microlensing in a lensed quasar can lead 
 to changes of continuum spectral
slopes and  line shapes in the spectra  of the quasar components. 
Comparing the 
spectra of different
components of the lensed quasar and the spectra of an image observed in
different epochs one can infer
the
presence of millilensing, microlensing and intrinsic variability.
Especially, microlensing can be used
for investigation of the unresolved broad line (BLR) and continuum 
emitting 
region
structure in active galactic nuclei (AGN). Therefore the spectroscopic 
monitoring of selected lensed quasars with 3D spectroscopy open new 
possibility for investigation of the  BLR structure in AGN. Here we 
discuss  observational effects that may be present during the BLR 
microlensing in 
the spectra
of lensed QSOs 

 \end{abstract}

\section{Introduction}
\label{sec:1}

Gravitational lensing is in general achromatic (the deflection angle of a
light ray does not depend on its wavelength); however, the
wavelength-dependent geometry of the different emission regions may
result in chromatic effects (see
Popovi\'c \& Chartas 2005, and references therein). Studies aimed at 
determining the 
influence
of microlensing on the spectra of lensed quasars (hereafter QSOs) 
need to account for the
complex structure of the QSO central emitting region. Since the sizes of
the emitting regions are wavelength-dependent, microlensing by stars in
a lens galaxy will lead to a wavelength-dependent magnification. The
geometries of the line and the continuum emission regions are in general
different and there may be a variety of geometries depending on the type
of AGN (i.e. spherical, disc-like, cylindrical, etc.). Observations and
modeling of microlensing of the broad-line region (BLR) of lensed QSOs
are promising, because the study of
the variations of the broad emission-line shapes in a microlensed QSO
image could constrain the size of the BLR and the continuum region.

Our knowledge of the inner structure of quasars 
is very limited and largely built on model calculations.
Continuum-line reverberation experiments with low-redshift QSOs
tell us that the broad-emission line region (BLR) is significantly
smaller than earlier assumed, and it is typically several light
days up to a light year across (e.g., Kaspi et al.\ 2000). It
means that the BLR radiation could be  significantly amplified due to
microlensing  by  (star-size) objects in an intervening galaxies 
(Abajas et al. 2002).
Hence, gravitational lensing can provide an additional method for
studying the inner structure high-redshift quasars for several
reasons:

(i) the extra flux magnification, from a few to 100 times,
provided by the lensing effect enables us to obtain high
signal-to-noise ratio (S/N) spectra of distant quasars with less
observing time;

(ii) the magnification of the spectra of the different images may
be chromatic (as was noted in Wambsganss \& Paczy\'nski 1991,
Wisotzki et al. 2003, Wucknitz et al. 2003, Popovi\'c \&
Chartas 2005) because of the line and continuum
emitting region are different in sizes and geometrically complex
and/or complex gravitational potential of lensing galaxy; (iia)
consequently, microlensing events lead to wavelength-dependent
magnifications of the continuum that can be used as indicators of
their presence (Wisotzki et al. 2003, Popovi\'c \& Chartas 2005);

(iii) gravitational microlensing can also change the shape of the
broad lines (see Popovi\'c et al. 2001, Abajas et al. 2002,
Popovi\'c \& Chartas 2005),  the deviation of the line profile
depends on the geometry of the BLR.

Finally, the monitoring  of lensed QSOs in order  to investigate
the effect of lensing on the spectra  can be useful
not only for constraining the unresolved structure of the central
regions of QSOs, but also for providing insight to the complex
structure of the lens galaxy.

\section{Structure of the central part  of AGNs and probability  of the 
BLR microlensing}

According to the standard model of AGNs, a QSO consists of a black
hole
surrounded by a (X-ray and optical) continuum emitting region probably
with an accretion disk geometry,
a broad line region (BLR) and a larger region, narrow line region (NLR) 
that can be resolved in several nearby
AGNs (e.g., Krolik 1999). The physics and structure of the NLR have been 
investigated using the observations of the region in nearby AGNs, while 
the physics and structure of the BLR cannot be investigated by direct 
observations.
Our knowledge of the inner structure of quasars
is very limited and largely built on model calculations.
Continuum-line reverberation experiments (in the UV/optical spectral band) 
with low-redshift QSOs
tell us that the broad-emission line region (BLR) is significantly
smaller than earlier assumed, and it is typically several light
days up to a light year across (e.g., Kaspi et al.\ 2000).

\begin{figure}[]
\includegraphics[width=6cm]{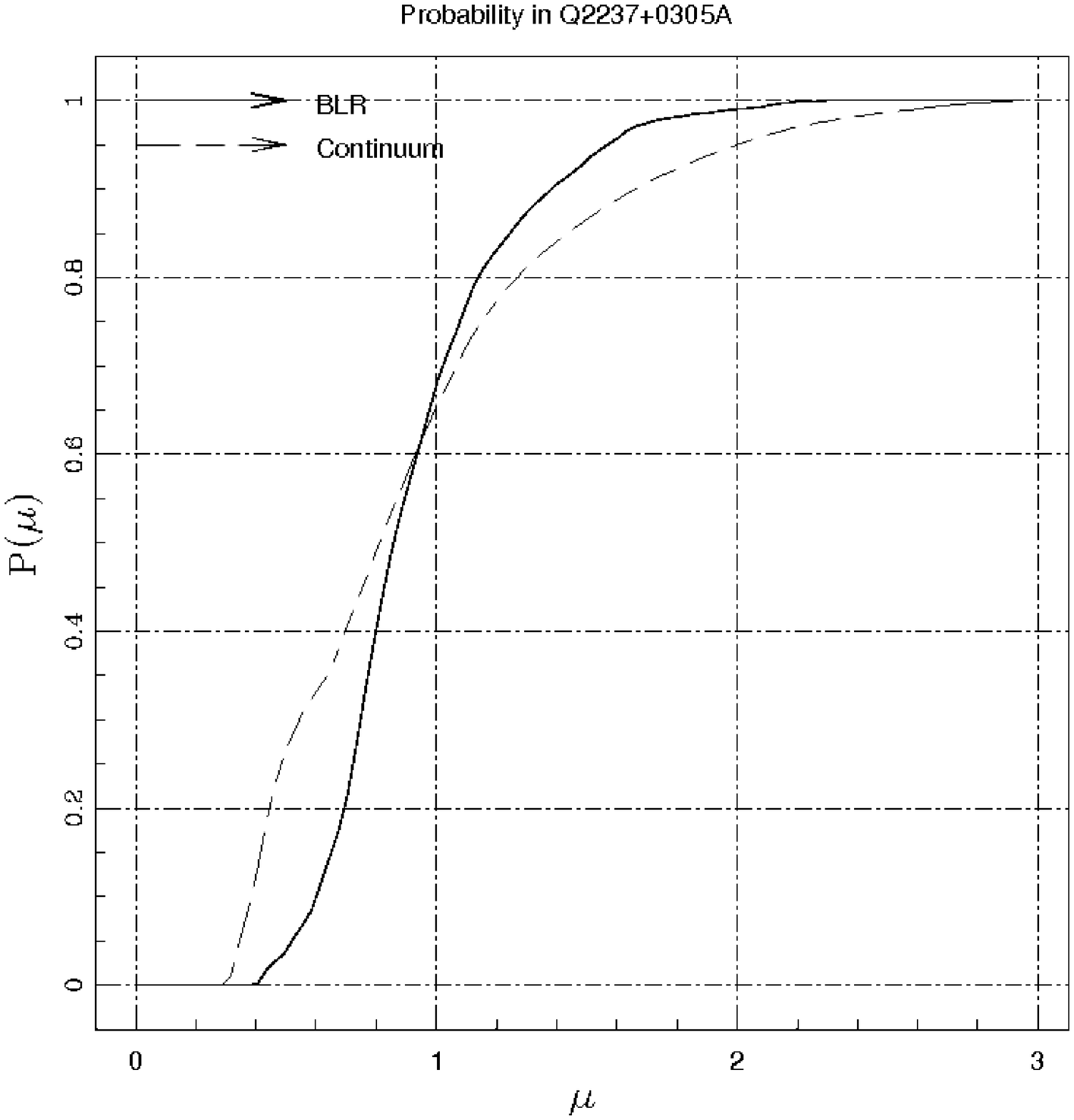}
\includegraphics[width=4cm]{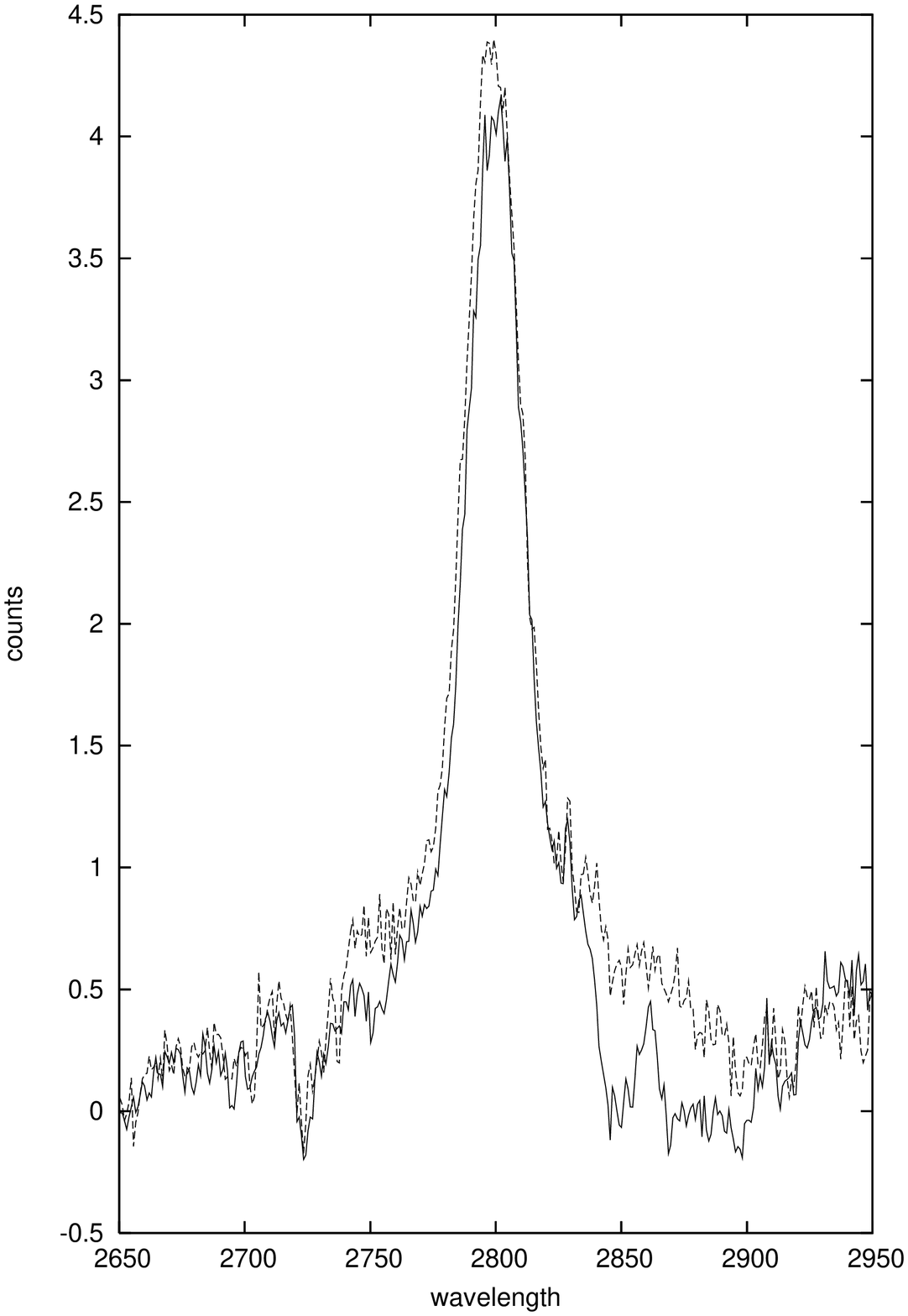}
 \caption{{\it Left}: Accumulate probability of the magnification of the
 BLR, in solid line, and the continuum, in dashed line (Abajas et
al. 2005). {\it Right}: Comparison of the line shape of Mg
II$\lambda$=2798 \AA\ of SDSS J0924+0219 observed at two epochs;
15/01/2005 -- solid line, and 01/02/2005 -- dashed line (Eigenbrod et 
al.
2005)} \label{f9} \end{figure}

Consequently, one can expect that the  magnification of the BLR (or a part 
of the BLR) and 
continuum 
emission due to microlensing. 
 In Fig. 1 (left), the cumulative probabilities for 
microlensing of the BLR and continuum as a function of the ERR (in units 
of the BLR dimensions) are given. As   one can see in Fig. 1 (left), there 
is a global 
correlation  between the BLR and continuum microlensing probabilities, i.e 
one can 
expect that the variation in the line profile of a QSO should be also 
seen as amplification in the continuum. We should note here that the 
emission of the BLR 
can be partially amplified due to microlensing, i.e. one part of the BLR 
can be affected by microlens that can result in amplification of the only 
one 
part of broad lines (see Fig 1, right)

\section{3D spectroscopy as tool for the BLR investigation}

A  systematic search for microlensing
signatures in the spectra of lensed quasars should be  performed with 
3D spectroscopy. One of examples is the Cosmological Monitoring of 
Gravitational Lenses (COSMOGRAIL, see Eigenbrod et al. 2005).
 To detect microlensing,  
simultaneous observations of spectra of images
of a lensed QSO at several epochs, preferably separated by the 
time-delay,
are needed (see Popovi\'c \& Chartas 2005).

\begin{figure}[]
\resizebox{\hsize}{!}{\includegraphics[]{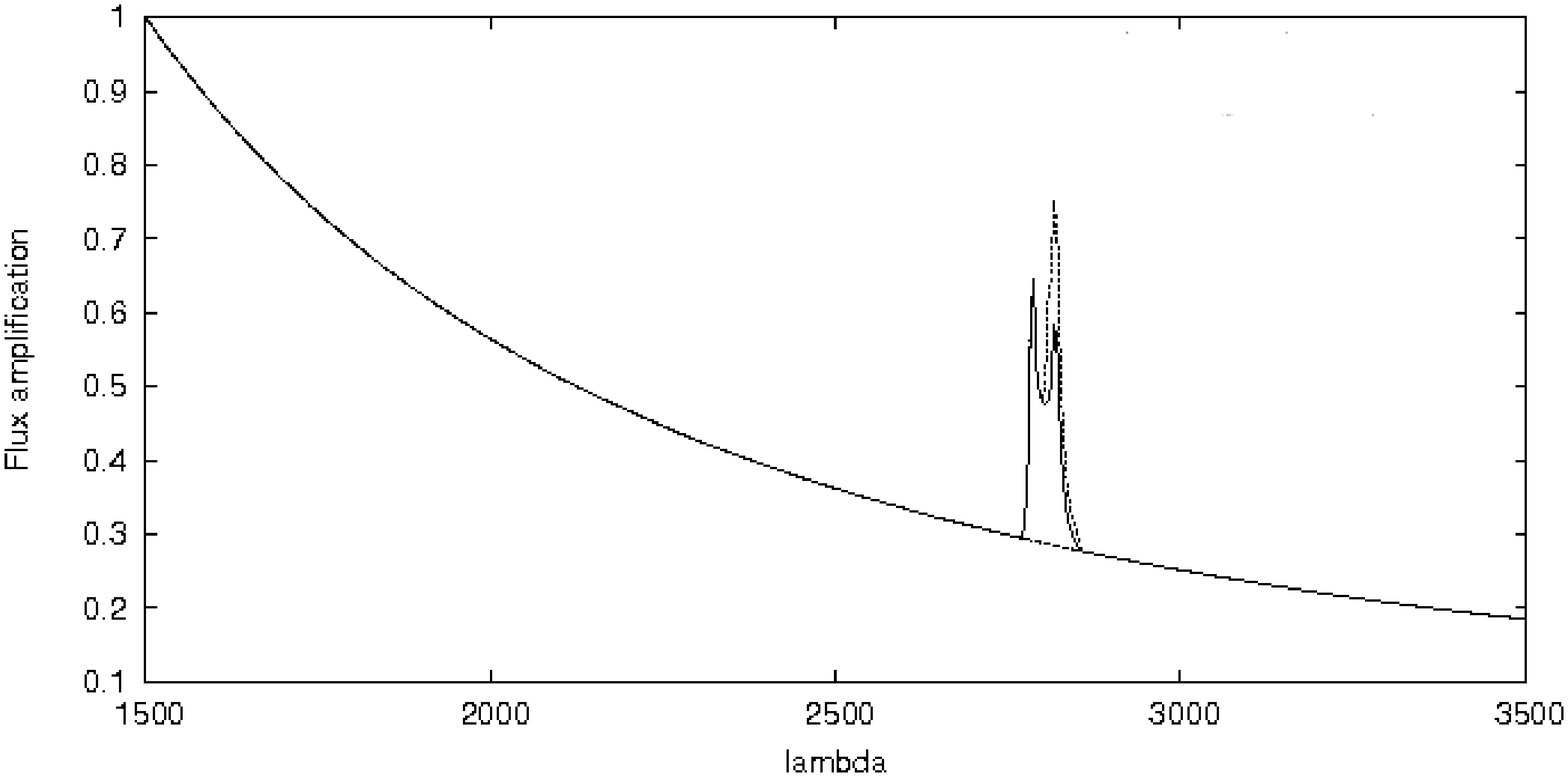}}
\resizebox{\hsize}{!}{\includegraphics[]{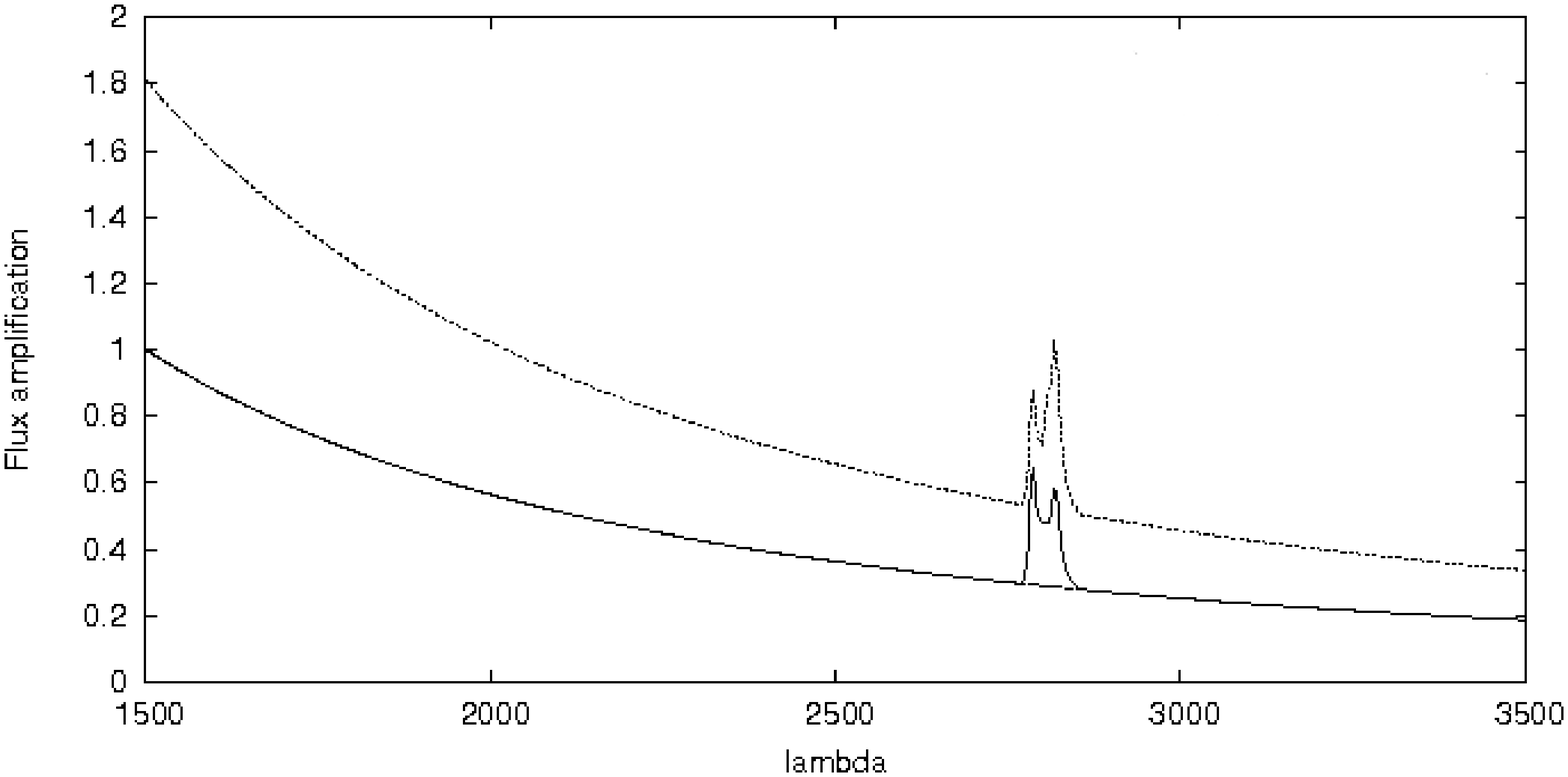}}
 \caption{The simulation of the variation in Mg II$\lambda=2798$\AA\ 
and the continuum between 1500 \AA\ and 3500 \AA\ due to microlensing by 
straight fold caustic. The microlensing may affect only line profile (top) 
as well as the line profile and the continuum (down). }
\label{f9}
\end{figure}

Taking into account the redshift of lensed QSOs it is needed to
obtain the spectra from 3500 to 9000 \ \AA\  (covering also, the
broad C IV, CIII] and Mg II lines which are emitted from the BLR
region) of a sample multi-imaged QSOs. Concerning the estimation of the 
BLR dimensions (see Kaspi et al. 2000), one should select a sample of 
lensed QSOs  where  the BLR  microlensing might be expected.  To find the 
possible microlensing one can apply the method given by
Popovi\'c \& Chartas (2005) comparing the spectra (in the
continuum and in the broad lines) of different components  in order
to detect the difference caused by microlensing or/and
millilensing.
 Using previous theoretical estimates of line shape
variations due to microlensing (Popovi\'c et al. 2001,
Abajas et. al. 2002, 2005, Popovi\'c \& Chartas 2005)  the observed 
spectra can 
be fitted with  the theoretical line profile assuming different geometries.
From this one will be able to  estimate the geometry and dimension
of the BLR. Also, comparing difference in amplification of the
continuum, C IV and Mg II lines one will be able to conclude about
differences between high and low ionized line emitting regions and
compare them with the size and geometry of the continuum emission
region.

\section{Microlensing of the BLR/continuum emission}

Let us discuss the expected variability in the line and continuum shapes 
due to microlensing. Recently, the broad line variability that may be 
caused by microlensing were observed by Richards et al. (2004) and 
Eigenbrod et al. (2005). Concerning the theoretical predictions (see 
Popovi\'c et al. 2001, Abajas et al. 2002) one can expect that different 
parts of a line can be amplified and that the line should change in 
intensity as well as in the shape during the microlensing event. Also, the 
continuum should vary (Lewis \& Ibata 2004) and  there is a 
correlation between the continuum and line amplification (see Fig. 1, 
left). Also,  it is expected that the continuum emission is wavelength 
dependent (Popovi\'c \& Chartas 2005). But, all of these effects might not 
be seen at the same time of a microlensing event. As an example in Fig. 2 
(top) we present simulation of microlensing of an emission from the 
accretion disk, taking that the continuum is coming from the inner part of 
the disk (from 50 Rg to 200 Rg) and the Mg II line from the outer part 
(from 200 Rg to 1200 Rg). The used model of the disk is the same as 
it is given in Popovi\'c 
et al. (2005), but with $i=14^\circ$ and for the UV radiation. In some 
cases 
the
continuum can remain constant, while the line can be amplified. It 
corresponds to 
the case where only the outer part  of the disk is microlensed (the part 
that emits the line). Such amplification of a line without amplification 
of the continuum can be seen only for a limited period of time. But 
during a complet microlensing event  the continuum should be also 
amplified (see Fig 2, down). Consequently, 
to 
register the microlensing presence, one should monitor lensed QSOs using 
3D 
spectroscopy in 
order to have observations of all images at the same time from different 
epochs.

\printindex
\end{document}